\documentclass[sigconf,table,xcdraw]{acmart} 
\usepackage{hyperref}  
\newcommand{\nop}[1]{}
\newcommand{\comm}[1]{\textcolor{red}{#1}} 

\usepackage{pifont} 
\usepackage{graphicx}     
\usepackage{bm}           
\usepackage{pifont}       
\usepackage{fontawesome5}

\usepackage{utfsym}           

\newcommand{\halfcheck}{{$\checkmark$\textsuperscript{{\kern-0.5em\tiny\scalebox{0.75}{\usym{2613}}}}}}

\usepackage{subfig} 
\usepackage{tcolorbox}    
\usepackage{tablefootnote}
\usepackage{threeparttable}
\usepackage{soul}

\tcbuselibrary{skins}     

\usepackage{multirow}     
\usepackage{longtable}    
\usepackage{diagbox}      
\usepackage{makecell}     
\usepackage{tabularx}     
\usepackage{booktabs}     

\usepackage[linesnumbered,ruled,vlined]{algorithm2e} 
\SetKw{Continue}{continue} 

\usepackage{algpseudocode}

\usepackage{listings} 

\definecolor{dkgreen}{rgb}{0,0.6,0}
\definecolor{gray}{rgb}{0.5,0.5,0.5}
\definecolor{mauve}{rgb}{0.58,0,0.82}

\setcopyright{none}  

\renewcommand\footnotetextcopyrightpermission[1]{}

\acmDOI{}               
\acmISBN{}              
\acmConference[]{} 
\acmYear{2025}          
\copyrightyear{}        
\acmPrice{}             

\begin{document}

\title{\textsf{AIMeter}: Measuring, Analyzing, and Visualizing Energy and Carbon Footprint of AI Workloads}

{
\author{Hongzhen Huang}
\affiliation{%
  \institution{The Hong Kong University of Science and Technology (Guangzhou)}
  \city{Guangzhou}
  \state{Guangdong}
  \country{China}
}
\email{hongzhenh326@gmail.com}

\author{Kunming Zhang}
\affiliation{%
  \institution{The Hong Kong University of Science and Technology (Guangzhou)}
  \city{Guangzhou}
  \state{Guangdong}
  \country{China}
}
\email{kzhang519@connect.hkust-gz.edu.cn}

\author{Hanlong Liao}
\affiliation{%
  \institution{National University of Defense Technology}
  \city{Changsha}
  \state{Hunan}
  \country{China}
}
\email{hanlongliao@nudt.edu.cn}

\author{Kui Wu}
\affiliation{%
  \institution{University of Victoria}
  \city{Victoria}
  \state{British Columbia}
  \country{Canada}
}
\email{wkui@uvic.ca}

\author{Guoming Tang}
\authornote{Corresponding author.}
\affiliation{%
  \institution{The Hong Kong University of Science and Technology (Guangzhou)}
  \city{Guangzhou}
  \state{Guangdong}
  \country{China}
}
\email{guomingtang@hkust-gz.edu.cn}

}

\begin{abstract}    
The rapid advancement of AI, particularly large language models (LLMs), has raised significant concerns about the energy use and carbon emissions associated with model training and inference. However, existing tools for measuring and reporting such impacts are often fragmented, lacking systematic metric integration and offering limited support for correlation analysis among them. This paper presents \textsf{AIMeter}, a comprehensive software toolkit for the measurement, analysis, and visualization of energy use, power draw, hardware performance, and carbon emissions across AI workloads. By seamlessly integrating with existing AI frameworks, \textsf{AIMeter} offers standardized reports and exports fine-grained time-series data to support benchmarking and reproducibility in a lightweight manner. It further enables in-depth correlation analysis between hardware metrics and model performance and thus facilitates bottleneck identification and performance enhancement.\nop{ Also, it incorporates multi-dimensional visualization capabilities that reveal dynamic relationships across metrics, thus improving interpretability.} By addressing critical limitations in existing tools, \textsf{AIMeter} encourages the research community to weigh environmental impact alongside raw performance of AI workloads and advances the shift toward more sustainable "Green AI" practices. The code is available at https://github.com/SusCom-Lab/AIMeter.
\end{abstract}

\maketitle  

\section{Introduction}

Artificial intelligence (AI) is advancing at an extraordinary pace, achieving state-of-the-art performance across domains such as natural language processing (NLP), computer vision (CV), and scientific computing. These breakthroughs are driven by increasingly large models, such as GPT-4, PaLM, and LLaMA, with hundreds of billions or even trillions of parameters trained on massive datasets using large-scale GPU clusters. However, this progress comes with growing environmental and economic costs. Training modern large language models (LLMs) can consume hundreds of megawatt-hours of electricity, with energy demands comparable to the annual usage of a small town~\cite{strubell2020energy,patterson2021carbon}. Meanwhile, LLM inference, especially in commercial deployments and operating at web scale, further compounds energy use over time. These trends raise pressing concerns about the sustainability of AI at scale, particularly in the face of global decarbonization goals and resource-constrained infrastructures.

Despite this, the AI research community remains largely focused on improving accuracy, scalability, and latency, with energy and carbon footprint treated as secondary or ignored entirely~\cite{schwartz2020green,henderson2020towards}. This omission carries practical, economic, and ethical consequences. On the economic front, a single LLM architecture search or fine-tuning experiment can incur thousands of dollars in cloud computing fees, much of which is attributable to energy-intensive GPU usage~\cite{lacoste2019quantifying}. 
\nop{For academic labs and early-stage startups, energy costs can be a major barrier to iteration and reproducibility.}
Environmentally, as AI becomes a larger component of digital infrastructure, its carbon footprint will increasingly come under scrutiny, especially as sustainability reporting frameworks (e.g., the EU’s Corporate Sustainability Reporting Directive~\cite{hummel2024overview}) mandate transparency in operational emissions. This makes it imperative to treat energy/carbon impact as a first-class metric in AI development and deployment.

To address the high energy costs and environmental impact of AI, an important research direction is the development of models capable of predicting energy consumption before task execution. This facilitates more proactive energy management and optimization. While some explorations have been conducted~\cite{chen2025much}, building reliable predictive models for varied AI tasks still requires a solid foundation. A key factor impacting the construction of such \nop{predictive}models is the need for comprehensive and standardized measurements. \nop{The complexity and resource demands of capturing detailed AI workload metrics result in a lack of comprehensive, standardized toolkits.}
\nop{which is inadequate as current approaches lack comprehensiveness and ease of use.}

Currently, although some tools are available for measuring and reporting energy and power consumption, they exhibit critical limitations. 

\begin{table*}[ht]
\centering
\scriptsize 
\renewcommand{\arraystretch}{1.3}

\begin{threeparttable}
\caption{Feature Comparison of Popular Tools}
\label{tab:tool_comparison}
\setlength{\tabcolsep}{1pt} 

\begin{tabular}{ >{\centering\arraybackslash}p{2.8cm} *{10}{>{\centering\arraybackslash}p{1.5cm}} }
\hline
\textbf{Tool / Feature} & 
\makecell{\textbf{Fine}\\\textbf{metrics}\hyperref[tabnote:a]{\textbf{\tnote{a}}}} & 
\makecell{\textbf{Coarse}\\\textbf{metrics}\hyperref[tabnote:b]{\textbf{\tnote{b}}}} & 
\makecell{\textbf{GPU}\\\textbf{power}} & 
\makecell{\textbf{CPU/RAM}\\\textbf{power}} & 
\makecell{\textbf{Carbon}\\\textbf{estimation}} & 
\makecell{\textbf{AI task}\\\textbf{sync.}} & 
\makecell{\textbf{Correlation}\\\textbf{analysis}} & 
\makecell{\textbf{Multi-dim}\\\textbf{visualization}} & 
\makecell{\textbf{Complexity\&}\\\textbf{overhead}} \\ 
\hline
\textbf{NVIDIA-SMI~\cite{nvidia-smi}} & Limited\hyperref[tabnote:c]{\tnote{c}} & Full & Full & Limited & Limited & Limited & Limited & Limited & Low \\
\textbf{DCGM~\cite{dcgm}} & Full & Full & Full & Limited & Limited & Limited & Limited & Limited & Low\\
\textbf{Nsight (nsys/ncu)~\cite{nsight-systems,nsight-compute}} & Full & Full & Full & Limited & Limited & Limited & Partial & Full & High\\ 
\textbf{PyJoules~\cite{pyjoules}} & Limited & Limited & Full & Full & Limited & Full & Limited & Limited & Medium\\
\textbf{CarbonTracker~\cite{anthony2020carbontracker}}& Limited & Limited & Full & Full & Full & Full & Limited & Limited & Medium\\
\textbf{CodeCarbon~\cite{benoit_courty_2024_11171501}}& Limited & Limited & Full & Full & Full & Full & Limited & Limited & Medium\\
\textbf{\textsf{AIMeter} (Proposed)} & Full & Full & Full & Full & Full & Full & Full & Full & Low\\
\hline
\end{tabular}

\begin{tablenotes} 
  \item[\phantomsection\label{tabnote:a}a] Refers to fine-grained metrics (down to the Streaming Multiprocessor (SM) and Tensor Core level), which are included in Appendix~\ref{sec:metrics_appendix}.
  \item[\phantomsection\label{tabnote:b}b] Refers to coarse-grained metrics, such as those collectible by \texttt{nvidia-smi} (e.g., GPU utilization, GPU memory utilization, temperature, etc.).
  \item[\phantomsection\label{tabnote:c}c] In the table, `Full` denotes that the feature is supported, `Partial` denotes partial support, and `Limited` denotes that the feature is unsupported.
\end{tablenotes}

\end{threeparttable}
\renewcommand{\arraystretch}{1} 
\end{table*}

\textbf{Fragmented measurement and limited carbon awareness.} While various tools, for example, NVIDIA-SMI, DCGM, and Nsight (Section~\ref{tools}), exist to collect metrics such as power consumption, energy usage, and hardware utilization for AI tasks, they often overlook carbon emissions and lack integration across these dimensions. This fragmentation hinders the generation of comprehensive, interpretable reports that reflect the true environmental cost of AI workloads. A unified framework that incorporates carbon estimation and supports standardized reporting is urgently needed.

\textbf{Complexity in usage and operational overhead.} Many similar tools are complex to deploy. The setup and configuration of these tools can be intricate. This process often requires specific dependencies or non-trivial integration efforts within diverse AI frameworks, and thus creates a barrier to entry for some users. \nop{Additionally, the monitoring process often consumes notable CPU or memory resources.}
Additionally, the monitoring process often has notable impacts, including high CPU usage, high memory usage, and significant time consumption. This overhead can interfere with the AI workload's performance.

\textbf{Insufficient in multi-dimensional visualization and correlation analysis.} While existing tools like CodeCarbon and PyJoules (Section~\ref{carbon-tools}) support tracking energy or carbon metrics, a major challenge lies in their inadequate visualization of multi-dimensional data such as energy and hardware state. This deficiency indirectly results in a lack of in-depth correlation analysis against hardware characteristics (e.g., GPU type, memory usage) or model performance indicators (e.g., latency, accuracy), hindering the observation of dynamic trends and limiting the fine-grained, data-driven optimization of AI workloads.

\nop{
\textbf{Insufficient in correlation analysis across metrics and system behavior.} Although existing tools such as CodeCarbon and PyJoules (Section~\ref{carbon-tools}) offer support for tracking energy or carbon metrics, they seldom facilitate in-depth correlation analysis with hardware characteristics (e.g., GPU type, memory usage) or model performance indicators (e.g., latency, accuracy). This lack of integrated analysis hampers the identification of efficiency bottlenecks and limits the ability to optimize AI workloads in a fine-grained, data-driven manner.

\textbf{Challenges in multi-dimensional metric visualization.}
Effectively presenting complex, multi-dimensional data, including energy use, carbon footprint, hardware state, and model behavior, remains a major challenge. Current tools offer limited support for intuitive, insightful visualizations, restricting researchers’ ability to observe dynamic trends and interpret system behavior. A more capable and visualization-aware toolkit is needed to close this gap.}

\nop{
\begin{itemize}
\item Generally, while many tools exist for acquiring metrics reflecting AI task energy costs, such as energy consumption, power consumption, and hardware information, there is a lack of focus on carbon emissions. There is also a lack of tools to integrate these metrics, thus preventing the formation of an effective, intuitive, and comprehensive report to demonstrate the energy cost of AI tasks.
\item Secondly, existing tools and research generally lack in-depth analysis of the correlations between these energy consumption/carbon emission metrics and specific hardware characteristics (such as GPU model, memory usage) as well as model performance (such as inference latency, accuracy). This lack of correlation analysis makes it difficult to understand and optimize AI tasks at a fine-grained level.
\item Third, concerning data presentation, how to effectively and intuitively visualize multi-dimensional data—including energy consumption, carbon emissions, hardware state, and model performance—to clearly reveal their complex relationships and trends, represents a significant challenge. Current shortcomings limit the effective demonstration and interpretation of research results.
\end{itemize}

\comm{GT: consider the following motivations: integration for a complete version of power/energy consumption and carbon footprint (for reporting purpose); correlation between those footprints and hardware/model performance (for analyzing purpose); multi-dimensional data visualizing (for demonstrating purpose).}

Motivated by these three key limitations, we decided to develop a dedicated tool for measuring, analyzing, reporting, and visualizing metrics including power consumption, energy consumption, hardware information, and carbon emissions of AI tasks.
}

By bridging existing research gaps and addressing corresponding challenges, we introduce \textsf{AIMeter} and make the following major contributions.

\begin{itemize}
    \item \textbf{Comprehensive Measurement:} \textsf{AIMeter} unifies real-time monitoring of energy, power, and hardware metrics with carbon emission estimation, offering synchronized tracking throughout AI workload execution. Upon task completion, \textsf{AIMeter} generates intuitive post-task reports and exports detailed time-series data, addressing key gaps in metric integration and carbon-aware reporting found in existing tools.
    
    \item \textbf{Ease of Use and Low Overhead:} \textsf{AIMeter} is simple to use and seamlessly integrates with various AI workflows; concurrently, it operates with minimal performance overhead (Section~\ref{experiments}). This addresses the challenge of a high barrier to entry, avoids impacting AI task performance, and ensures reliable long-term measurement.
    
    \item \textbf{Multi-Dimensional Visualization and Correlation Analysis:} \textsf{AIMeter} implements multi-dimensional visualization capabilities, effectively demonstrating how metrics such as energy consumption, carbon emissions, and hardware characteristics evolve over time and relate to one another. Through these tailored visualizations, correlation analysis across classified metrics spanning compute, memory, and communication dimensions is enabled. This bridges a key gap in existing tools by supporting multi-dimensional, workload-aware efficiency analysis.
    
    \nop{
    \item \textbf{In-depth Correlation Analysis:} \textsf{AIMeter} enables detailed investigation of how energy consumption, power usage, and carbon emissions correlate with hardware behavior and model performance. By coupling fine-grained time-series data with classified metrics spanning compute, memory, and communication, it allows researchers to identify bottlenecks and assess their impact on energy and environmental costs. This bridges a key gap in existing tools by supporting multi-dimensional, workload-aware efficiency analysis.
    \item \textbf{Multi-Dimensional Visualization:} \textsf{AIMeter} incorporates tailored visualization capabilities to effectively demonstrate the complex interplay among metrics such as energy consumption, carbon emissions, and hardware characteristics. These include real-time monitoring dashboards, post-hoc analytical plots, structured tables, and summary reports designed to intuitively reveal dynamic trends and cross-metric correlations. The visualizations help uncover nuanced behaviors, thereby enhancing interpretability and supporting data-driven optimization.}
\end{itemize}

\nop{\textbf{Comprehensive Reporting:} We have organically integrated multiple metric collection interfaces, enabling real-time, synchronous measuring of AI tasks' energy consumption, power consumption, and hardware metrics, while innovatively introducing carbon emission calculations. Consequently, after task completion, \textsf{AIMeter} can generate an effective, intuitive, and comprehensive report that not only summarizes key information like task duration, total energy consumption, and estimated carbon emissions, but also saves detailed time-series data as CSV files or stores it in a database. This overcomes the shortcomings of existing tools in metric integration and carbon emission reporting.

\textsf{AIMeter} enables in-depth analysis of the correlations between energy consumption, power consumption, carbon emissions, and specific hardware characteristics as well as model performance. By providing metric data tightly coupled with the time series, researchers can explore, for example, how specific GPU models or memory usage patterns impact power consumption and carbon emissions. We have classified different types of metrics (compute, memory, and communication) to help identify bottlenecks during the AI task process and link these bottlenecks with indicators such as instantaneous power consumption, thereby enabling a more fine-grained understanding and optimization of AI tasks. This fills a gap in current research regarding such multi-metric correlation analysis.

To address the challenge of effectively demonstrating the complex relationships among multi-dimensional metrics such as energy consumption, carbon emissions, and hardware attributes, we have developed specialized visualization functions. These functions provide effective and intuitive multi-dimensional data visualization solutions, including reports, tables, real-time monitoring, and post-hoc analysis plotting. The design emphasizes clearly revealing the complex interrelationships and dynamic trends among various metrics, thus greatly serving the demonstration and interpretation of results. For example, the visualizations generated by this tool can clearly demonstrate the significant dynamic changes in power consumption and GPU metrics during different stages of LLM inference.

\textbf{Benchmark for Energy:} We hope that through \textsf{AIMeter}, a measurement standard focused on AI energy consumption can be established, allowing for better evaluation of the contributions of different research efforts in the field of Green AI.}

Ultimately, our aim is not only to provide an effective and comprehensive toolkit for benchmarking energy, power, carbon emissions, and hardware metrics of AI workloads, but also to promote deeper awareness of the environmental costs embedded in AI research and development. 
\nop{We envision \textsf{AIMeter} as a bridge toward more sustainable AI practices—supporting the transition from performance-centric "Red AI" to energy-conscious, environmentally responsible ``Green AI''~\cite{schwartz2020green}.}

\section{Related Work}

In this section, we review existing tools (shown in Table~\ref{tab:tool_comparison}) capable of measuring energy, power consumption, hardware metrics, and carbon emissions related to AI workloads.

\subsection{Hardware Monitoring Interfaces and Libraries}

\label{tools}
NVIDIA provides Nsight Systems (nsys)~\cite{nsight-systems} and Nsight Compute (ncu)~\cite{nsight-compute}, which possess deep performance profiling capabilities with fine granularity, effective for identifying computational bottlenecks. However, their focus is on performance debugging, and their high sampling precision (especially ncu) incurs high operational overhead, making them unsuitable as low-interference, long-duration measurement tools.

NVIDIA-SMI~\cite{nvidia-smi} is a command-line utility provided by NVIDIA for managing and monitoring GPU devices; its overhead is very low, but it only provides coarse-grained metric measurements, such as overall GPU power consumption and utilization. NVIDIA Data Center GPU Manager (DCGM)~\cite{dcgm} offers finer granularity, capable of acquiring a wide range of relatively fine-grained metrics, including Tensor Core and SM activity status, suitable for large-scale cluster measurements. However, integrating these tools into a unified measurement framework precisely synchronized with the AI task execution process and including energy consumption and carbon emission estimation requires researchers to exert additional effort for secondary development and metric integration (such as Intel RAPL~\cite{rapl}). 


\begin{figure*}[htbp]
    \centering
    \includegraphics[width=0.9\linewidth]{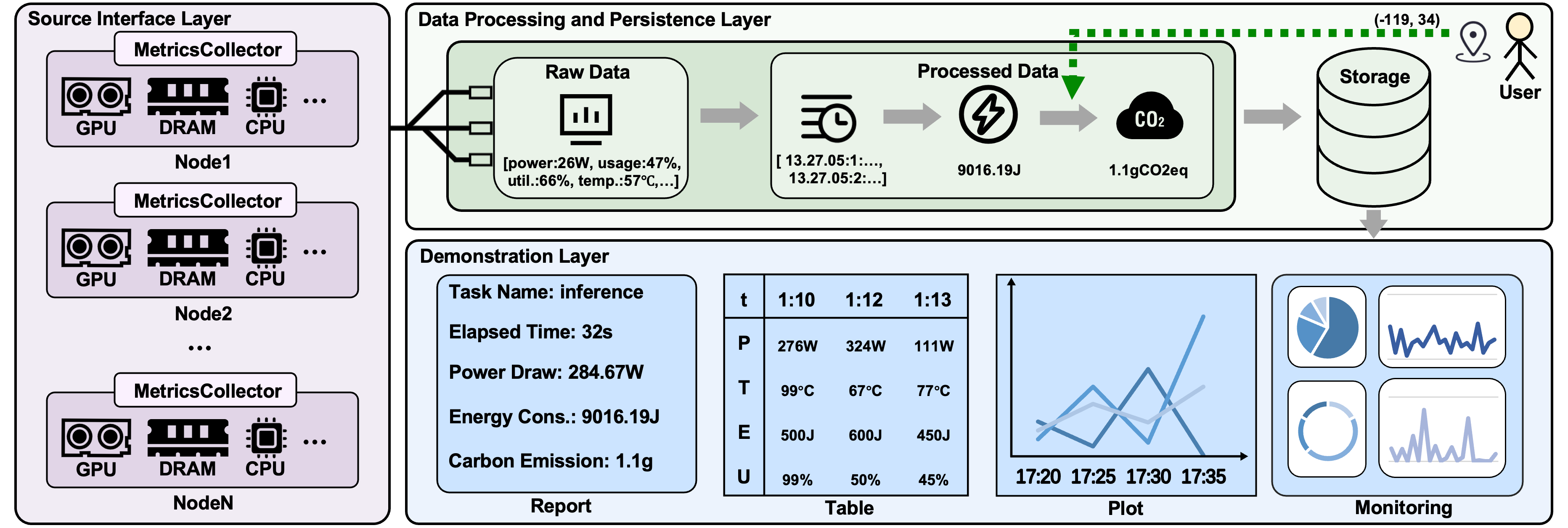 } 
    \caption{The architecture of \textsf{AIMeter}.}
    \label{fig:architecture}
\end{figure*}

\subsection{Energy Measurement Frameworks and Tools}
\label{carbon-tools}

PyJoules~\cite{pyjoules} is a Python library that utilizes interfaces like Intel RAPL~\cite{rapl} and NVIDIA NVML~\cite{nvml} to retrieve hardware energy consumption data. CodeCarbon~\cite{benoit_courty_2024_11171501} is a relatively mature tool specifically focused on estimating the carbon emissions produced during code execution. It operates by measuring the energy consumption of key hardware components (such as CPU, GPU, and RAM) and then applying location-based carbon intensity factors to convert this energy consumption into a carbon footprint. CarbonTracker~\cite{anthony2020carbontracker} is also a tool with similar functionality to CodeCarbon. Besides these, there are also system-level measurement and cloud platform tools, but their granularity is generally coarser.

Existing energy and carbon measurement tools exhibit several shortcomings. Firstly, these tools typically report only total or average values, which restricts their capability for deep correlation analysis and makes it difficult to attribute specific energy or power consumption changes to underlying hardware events during AI tasks. Secondly, their visualization capabilities tend to concentrate on single metrics, such as carbon emissions, often lacking the interactive, multi-dimensional perspectives needed to easily interpret complex system behaviors.

\nop{It can be seen that existing tools provide measurements for various parts of AI workloads, but the fact that each part is fragmented and independent is precisely a key challenge. How to comprehensively measure and evaluate the energy consumption and carbon emissions of an AI task, and understand the complex relationship between hardware behavior and model performance, often requires researchers to spend significant practical effort and time synchronizing heterogeneous data sources, writing analysis scripts, and implementing visualizations. This high barrier hinders in-depth research into AI sustainability.}

\section{Architecture and Design}

\nop{
\subsection{Architecture Overview}  
Fig.~\ref{fig:architecture} shows the architecture of \textsf{AIMeter}, which is fundamentally divided into three layers:
\begin{itemize}
    \item \textbf{Source Interface Layer:} This layer is primarily responsible for collecting power consumption, energy consumption, and hardware metric data using interfaces provided by the hardware (e.g., Intel RAPL, NVIDIA SMI, DCGM).
    \item \textbf{Data Processing and Persistence Layer:} This layer handles the raw data post-collection. Its responsibilities include transforming certain metrics into more intuitive formats and ensuring tight coupling between All Metrics and their corresponding timestamps in the time series. Once processed and time-aligned, the data is stored persistently (e.g., in files or a database) for later use in presentation and further analysis.
    \item \textbf{Demonstration Layer:} Upon completion of the AI task, this layer presents the processed results to the user. It supports multiple output formats.
\end{itemize}
Through this three-layer architecture, \textsf{AIMeter} systematically measures, analyzes and visualizes energy consumption, power consumption, carbon emissions and other relevant hardware metrics throughout the execution of AI tasks.

In the following sections, we describe the design and implementation of \textsf{AIMeter} at different levels of granularity.
}

Fig.~\ref{fig:architecture} shows the architecture of \textsf{AIMeter}, which is fundamentally divided into i) source interface layer, ii) data processing and persistence layer, and iii) demonstration layer. Through this three-layer architecture, \textsf{AIMeter} systematically measures, analyzes and visualizes energy consumption, power consumption, carbon emissions and other relevant hardware metrics throughout the execution of AI tasks.

\subsection{Source Interface Layer}

As mentioned in the previous section, metric data collection primarily relies on interfaces provided by the hardware. Based on user configuration, \textsf{AIMeter} dynamically selects the metrics to sample and can support up to 26 distinct metrics. These include GPU power consumption, SM active, among others.

Furthermore, a key design goal is to achieve the highest sampling frequency permitted by the underlying hardware interfaces. This high temporal resolution allows researchers to observe subtle variations in energy consumption and power draw across different fine-grained stages within AI task execution.

Given that different hardware components expose metrics through distinct interfaces, and even a single component like an NVIDIA GPU might offer data via multiple mechanisms (e.g., NVML/SMI and DCGM), we employ a parallelized collection strategy. This is handled within \textsf{AIMeter} by our \textit{MetricsCollector} class, which is responsible for the statistical metric collection. Where appropriate (primarily for GPU metrics), this allows \textsf{AIMeter} to query interfaces like those underlying SMI and DCGM concurrently. This approach maximizes the achievable sampling rate to approximately 0.1s-0.2s per sample, a frequency no lower than that of pynvml~\cite{pynvml}.

\nop{
For comparison, we consider pynvml~\cite{pynvml} as a baseline. Pynvml provides Python bindings for NVML~\cite{nvml}, offering a way to access GPU metrics programmatically. Given its common usage and similar underlying library access (for some metrics), we selected pynvml as our baseline for comparative performance evaluation regarding sampling frequency. Experiments were conducted across three distinct server setups. The experimental results demonstrate the high sampling frequency capability of our toolkit.

\begin{table}[ht]
\centering
\small 
\renewcommand{\arraystretch}{1.3} 
\caption{Sampling Frequency Comparison}
\label{tab:time_consumption_wrap}
\begin{tabular}{ >{\centering\arraybackslash}p{2.75cm} c c c }
\hline
\textbf{Time consumption} & \textbf{A800} & \textbf{4xA6000} & \textbf{8xRTX4090} \\
\hline
\textbf{Pynvml} & 186.412ms & 104.730ms & 857.701ms \\
\textbf{\textsf{AIMeter}} & 33.759ms & 97.657ms & 157.615ms \\
\hline
\end{tabular}
\renewcommand{\arraystretch}{1} 
\end{table}

Table~\ref{tab:time_consumption_wrap} displays the experimental results. On each of the three server platforms, we measure the achievable sampling frequency when collecting basic metrics (indicators like GPU power consumption, GPU utilization, memory utilization, GPU temperature, and clock speeds):

The results clearly show that, for the same metric collection profile on the same hardware, \textsf{AIMeter} consistently achieves higher sampling frequencies than the pynvml baseline across all tested platforms. The performance gain is especially significant on the server equipped with A800 GPU, where our tool demonstrated sampling rates approximately 6 times higher than the baseline.
}

\subsection{Data Processing and Persistence Layer}

After collected from the hardware, the raw data go through the following two-stage converting processes.

\textbf{Stage-1:} Normalization involves converting cumulative energy metrics (e.g., Joules) obtained for CPU and DRAM via interfaces into average power consumption (e.g., Watts) over the sampling interval. This unit regularization simplifies subsequent analysis and offers a more intuitive view of resource usage.

\textbf{Stage-2:} Energy consumption data is converted into estimated carbon emissions. This conversion is critical for meaningful environmental impact reporting, as grid carbon intensity factors vary significantly across both geography and time. 
\nop{Simply reporting energy usage fails to capture the actual environmental footprint.} 
As demonstrated by article~\cite{wu2025carbonedge}, substantial real-time carbon intensity differences exist even at mesoscales (tens to hundreds of km), with simultaneous variations reaching factors of 7.9× (Western US) and 19.5× (Central Europe) between adjacent areas. Furthermore, significant temporal fluctuations occur, including intraday and seasonal variations. This evidence strongly emphasizes the dynamic nature of carbon intensity. 
\nop{and the necessity of using accurate, context-specific (time and location) factors for reliable emissions estimation.}

This conversion process raises two key questions. The first concerns the distinction between embodied and operational carbon~\cite{li2024carbon}.

\begin{itemize}

    \item \textbf{Embodied carbon} encompasses the emissions from manufacturing, transport, etc., representing a one-time upfront cost associated with the hardware itself.
    
    \item \textbf{Operational carbon} stems from the energy consumed during the hardware's use phase. \nop{(i.e., electricity generation emissions).}
    
\end{itemize}

While embodied carbon is crucial for comparing different hardware solutions or for full Life Cycle Assessments (LCA), \textsf{AIMeter}'s primary focus is on quantifying the direct impact of specific AI tasks (training, inference, etc.) during their execution. Consequently, we deliberately exclude embodied carbon and measure only the operational carbon generated during task execution, using the following rule:
\begin{equation}
Carbon = Energy \times Intensity.
\end{equation}

The second consideration is the type of carbon intensity factor used. \emph{Average intensity} reflects the grid's overall generation mix and is suited for broad estimations. \emph{Marginal intensity}, however, represents the emissions from the power source(s) activated to meet the additional load imposed by the task. Since this marginal generation often comes from more carbon-intensive sources (like fossil fuel peaker plants), its intensity is typically higher and more accurately reflects the direct environmental consequence of running the AI task. To provide a clearer picture of the incremental impact and environmental cost, \textsf{AIMeter} uses marginal carbon intensity for its estimations, with the necessary intensity factors being retrieved from services like Electricity Maps~\cite{electricity_maps} and WattTime~\cite{Watttime} based on geographical coordinates.

\nop{To provide a clearer picture of the incremental impact and environmental cost, \textsf{AIMeter} uses marginal carbon intensity for its estimations. 

\textsf{AIMeter} uses geographical coordinates to get carbon intensity factors from services like Electricity Maps~\cite{electricity_maps} and WattTime~\cite{Watttime}.}

\nop{To obtain the intensity factors, \textsf{AIMeter} leverages data from services like Electricity Maps~\cite{electricity_maps} and WattTime~\cite{Watttime}, using user-provided geographical coordinates (latitude and longitude) to fetch the relevant real-time or regional marginal carbon intensity. }

\nop{The calculated carbon emissions are then included in the final report.}

Finally, all processed and normalized data, including these carbon estimates, are strictly time-aligned within the data series before being persistently stored in files or a database.

\subsection{Demonstration Layer}

The demonstration layer retrieves the processed and stored data, potentially applying final transformations for display, and presents it to the user. The goal is to provide multi-faceted data visualization options, enabling intuitive and comprehensive analysis. Key demonstration methods include:

\begin{itemize}
    
    \item \textbf{Execution Report:} After the task finishes, a summary report is generated.

    \item \textbf{Tabular Data:} Provides detailed, time-stamped metric information in a table format, suitable for export (e.g., CSV) or direct inspection.

    \item \textbf{Plotting:} Users can generate visualizations, typically line graphs, from the stored time-series data to observe the behavior of metrics over the task's duration.

    \item \textbf{Real-time Monitoring Dashboard:} We also integrate \textsf{AIMeter} with visualization tools (e.g., Grafana~\cite{Grafana}) for live monitoring during task execution.
    
\end{itemize}

These multiple visualization approaches (referring to Appendix~\ref{demonstration} for examples) combine to offer a more detailed and comprehensive view of the collected data.

\section{Case Study} 
\label{experiments}
\vspace{1em} 
\begin{lstlisting}[basicstyle=\footnotesize]
from AIMeter import monitor
try:
    monitor.start(sampling_interval = 0.1, ...)
    # Llama2-7b inference (sleep 15s between Prefill and Decode)
finally:
    monitor.stop()
\end{lstlisting}

Utilizing \textsf{AIMeter} as demonstrated in the code snippet above, we conducted an LLM inference experiment using the Llama2-7b model on a server with NVIDIA A800 GPU,\nop{To facilitate detailed observation, we intentionally used a large input sequence length, specifically an \textit{input\_size} of 40000 with a \textit{batch\_size} of 2, and limited the \textit{output} to 100.} and sampled All Metrics (see Appendix~\ref{sec:metrics_appendix} for the full list) at a sampling interval of $0.1$ second. With the help of \textsf{AIMeter}, this experiment yields findings with the following insights and implications. \nop{The underlying phenomena were consistently reproduced on other server configurations.}

\subsection{Correlation Analysis on Phase Dynamics}
\label{subsec:inference_dynamics} 

\begin{figure}[htbp]
    \centering
    \includegraphics[width=0.49\linewidth]{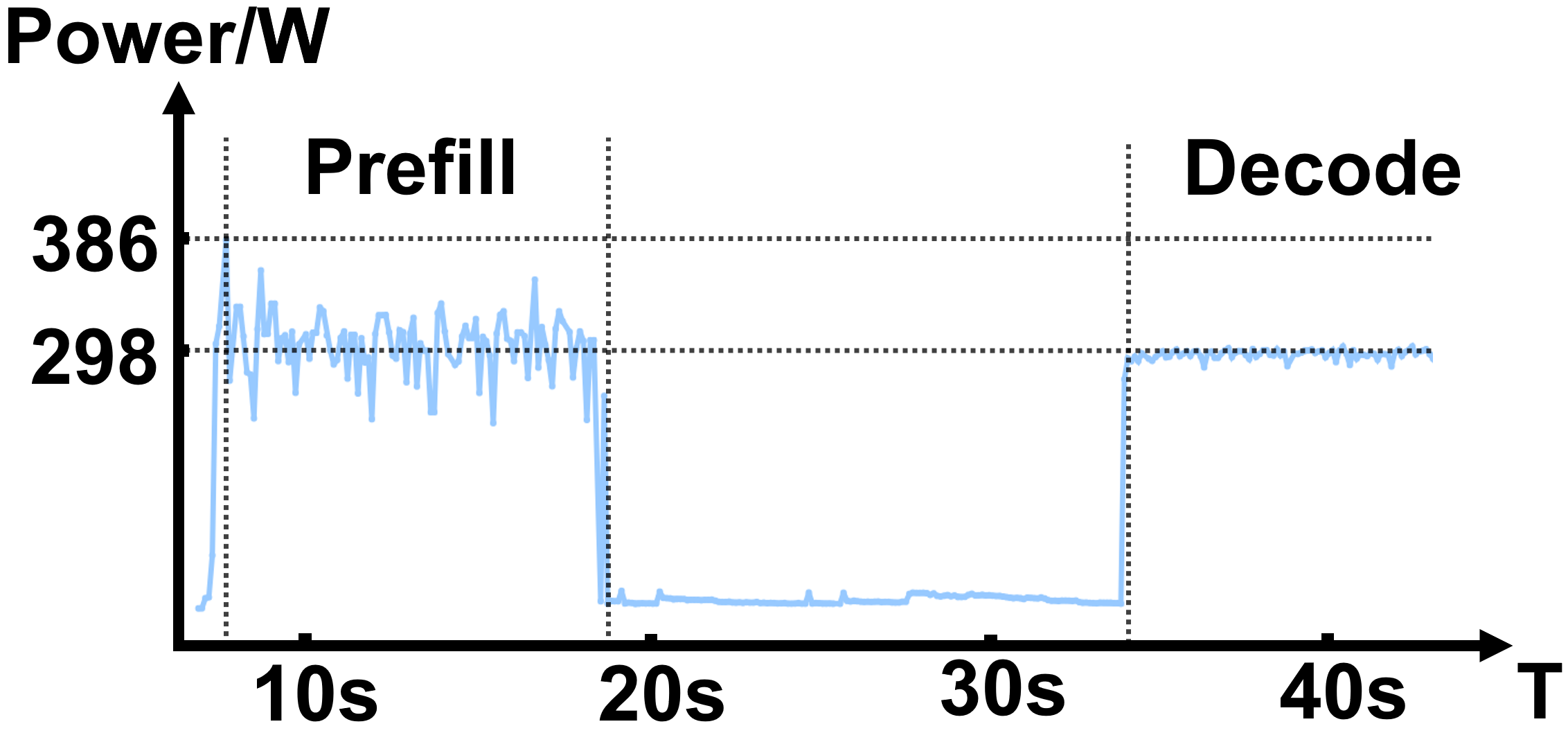}
    \includegraphics[width=0.47\linewidth]{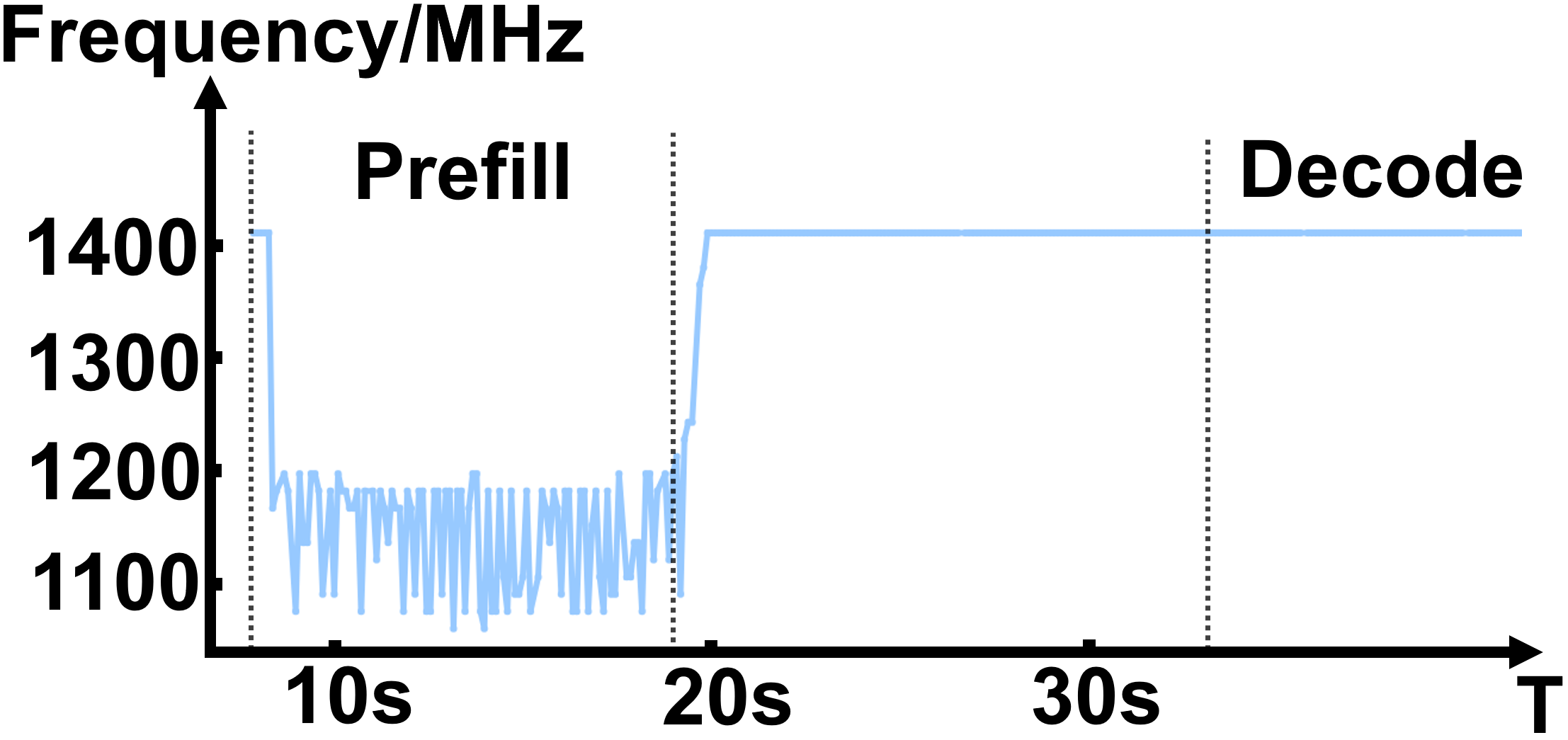}
    \caption{Power and frequency during inference phase.}
    \label{fig:power and frequency}
\end{figure}

For inference tasks, the LLM sequentially goes through the \emph{Prefill phase} and \emph{Decode phase} in generating the output (tokens). Observations from Fig.~\ref{fig:power and frequency} indicate that the Prefill phase has significantly higher peak power consumption (exceeding the Decode phase by nearly 90W) and exhibits larger power fluctuations. Consequently, the Prefill phase is the primary contributor to peak power demand, making optimization of its computational efficiency (e.g., using techniques like FlashAttention) crucial for reducing overall peak power. GPU frequency patterns mirror this, with the Prefill phase operating at lower and more variable frequencies, while the Decode phase sustains higher and more stable frequencies.

\begin{figure}[b]
    \centering
    \includegraphics[width=1\linewidth]{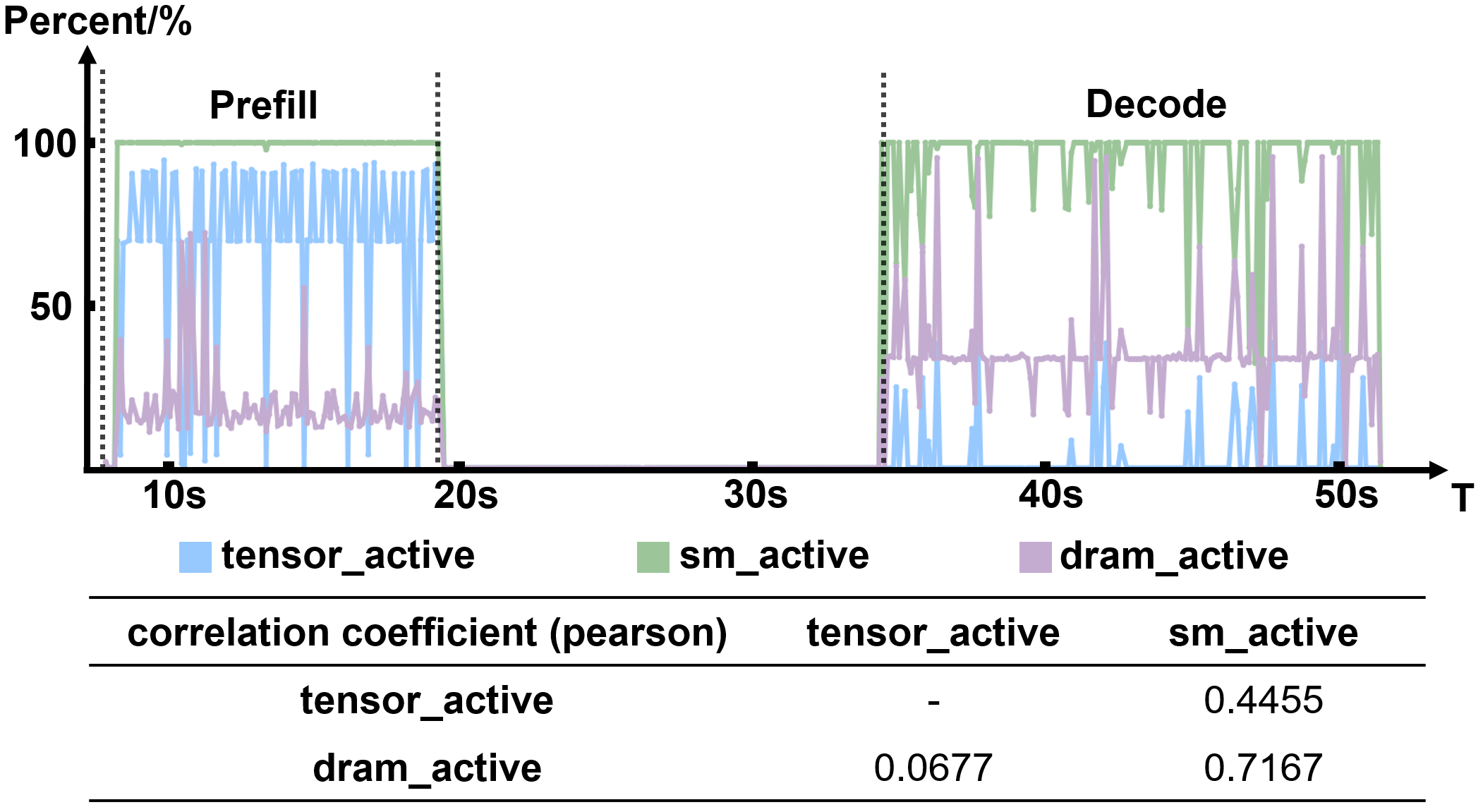} 
    \caption{Shifting bottlenecks between phases.}
    \label{fig:bottleneck}
\end{figure}

The two phases also differ in their primary performance bottlenecks, as shown in Fig.~\ref{fig:bottleneck}. The Prefill phase is predominantly compute-bound, whereas the Decode phase is mainly memory-bound.\nop{These differences necessitate distinct optimization strategies for each phase.} The underlying reasons stem from their operational nature: Prefill compute-intensively processes the full input in parallel, whereas Decode sequentially generates tokens, frequently accessing the large KV cache for smaller computations to predict the next token.

\nop{Prefill involves compute-intensive parallel processing of the entire input sequence (e.g., large matrix multiplications); conversely, the Decode phase sequentially generates tokens, leading to frequent memory access for the increasingly large Key-Value (KV) cache to perform relatively smaller computations for predicting the next token.}

\begin{figure}[t]
    \centering
    \includegraphics[width=0.8\linewidth]{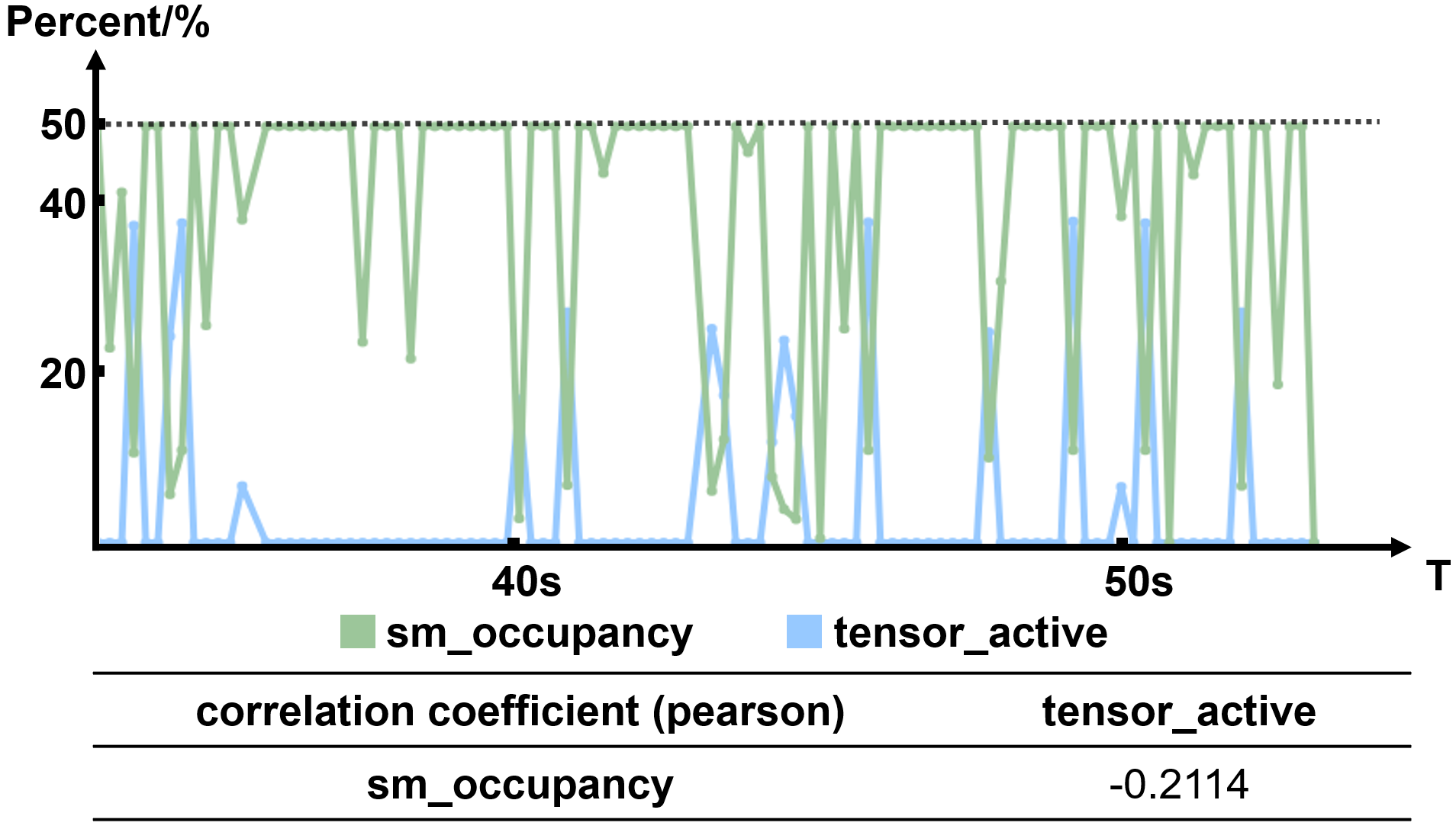} 
    \caption{Occupancy and active during decode phase.}
    \label{fig:mywidefigurelabe}
\end{figure}

Further analysis of GPU utilization during the inference process (Fig.~\ref{fig:mywidefigurelabe}) highlights a strong inverse correlation between \emph{SM occupancy} and \emph{Tensor active} percentages.\nop{These opposing trends suggest potential trade-offs or shifting dominant resource demands.} A reasonable explanation is that the GPU switches its main effort during inference: when it is doing heavy math, the specialized Tensor Cores are busy and SMs mostly assist; when it is doing other general tasks, the SMs are busy ones and the Tensor Cores are quieter. As previously analyzed and validated in Fig.~\ref{fig:mywidefigurelabe}, the Decode phase is not compute-intensive and thus the Tensor active often remains low during this phase.

\begin{figure}[htbp]
    \centering
    \includegraphics[width=0.7\linewidth]{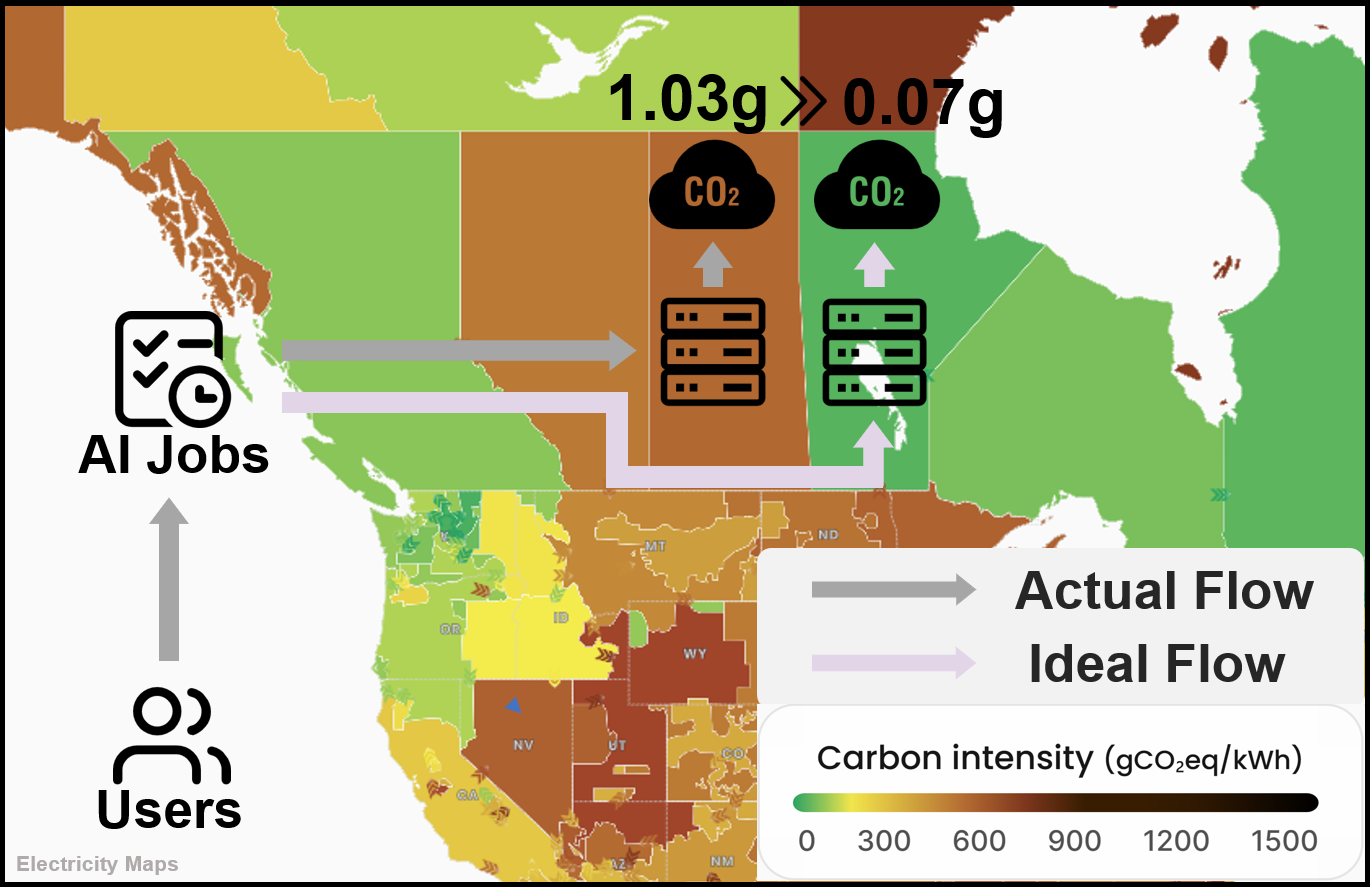} 
    \caption{Carbon emissions from the same AI task in two neighboring provinces of Canada, respectively.}
    \label{fig:carbon}
\end{figure}

\subsection{Analysis on Carbon Emission Estimation}

In the aforementioned experiment, the report generated by \textsf{AIMeter} indicated that the \nop{LLM inference}task consumed 7485 Joules (2.08 $Watts\cdot{hour}$) of energy. However, as we have mentioned, such energy consumption figure does not fully reflect the task's actual environmental impact, which motivated our introduction of the carbon emissions metric.

To demonstrate the decisive role of geographical location, we performed a comparative analysis: assuming that this task was executed in the Canadian provinces of Saskatchewan and Manitoba respectively, Fig.~\ref{fig:carbon} shows that the identical AI task could yield a difference in generated carbon emissions, 1.03g vs 0.07g, approaching 15-fold between these neighboring provinces. 

This clearly indicates that geographical location is a critical factor when assessing the environmental footprint of AI tasks. This finding aligns with research such as~\cite{nguyen2024towards}, who also demonstrate significant variations in operational carbon emissions for identical LLM tasks when executed in regions with differing grid carbon intensities, such as Québec, California, and the US PACE territories. Indeed, converting energy consumption to carbon emissions using location-specific carbon intensity factors, as employed in our analysis and detailed by~\cite{nguyen2024towards}, is crucial for accurate environmental assessments.

\subsection{Overhead Caused by \textsf{AIMeter}}

We evaluated the system overhead of \textsf{AIMeter} using \texttt{pidstat} across two control groups. The first group measured the inherent overhead of \textsf{AIMeter} by running it on an otherwise idle system with a 0.1-second sampling interval. The second group replicated the experiment from Section~\ref{experiments}, but with the modification that we removed the 15-second sleep time between the prefill and decode phases. In this setup, we evaluated the impact of \textsf{AIMeter} with sampling intervals of 0.1, 0.5, 1, and 5 seconds. The results are presented in Table~\ref{tab:overhead}.


\begin{table}[ht]
\centering
\footnotesize
\setlength{\tabcolsep}{2pt} 
\renewcommand{\arraystretch}{1.3} 
\captionsetup{skip=7pt}
\caption{Overhead Caused by \textsf{AIMeter} (AIM)}
\label{tab:overhead}
\begin{tabular}{ >{\centering\arraybackslash}p{2.1cm} c c c c c} 
\hline
\textbf{\makecell{Overhead}} & 
\textbf{\makecell{Elapsed\\time}} & 
\textbf{\makecell{Time\\Overhead}} & 
\textbf{\makecell{Time\\/sample}} &
\textbf{\makecell{CPU\\util.}} & 
\textbf{\makecell{Memory\\cons.}} \\

\hline
\textbf{Baseline} & 15s & --- & --- & 0\% &121.21MB  \\
\textbf{AIM(0.1) only} & 15.05s & 0.33\% \,$\uparrow$ & 0.37ms & 2.23\% & 121.29MB \\
\textbf{Exp w/o AIM} & 29.15s & --- & --- & 99.1\% & 1411.37MB \\
\textbf{Exp w AIM(5)} & 29.39s & 0.82\% \,$\uparrow$ & 38.51ms & 97.21\% & 1388.59MB \\ 
\textbf{Exp w AIM(1)} & 29.53s & 1.30\% \,$\uparrow$ & 12.46ms & 97.67\% & 1398.04MB \\ 
\textbf{Exp w AIM(0.5)} & 29.71s & 1.92\% \,$\uparrow$ & 9.28ms & 98.24\% & 1398.84MB \\ 
\textbf{Exp w AIM(0.1)} & 30.54s & 4.77\% \,$\uparrow$ & 7.2ms & 98.97\% & 1403.62MB \\ 
\hline
\end{tabular}
\renewcommand{\arraystretch}{1} 
\end{table}

\nop{Among these, the calculation method for time consumption per sample is to use the completion time of the experiment without \textsf{AIMeter} as a baseline, to calculate how much additional time overhead is brought by recording each sample when \textsf{AIMeter} is present.}


The data in the table indicates that the overhead caused by \textsf{AIMeter} is low. Besides, it can be seen that the higher the sampling frequency, the higher the overhead, but the time consumption per sample is smaller. Regarding CPU utilization, the highest usage is observed when \textsf{AIMeter} is absent, as the tool itself consumes a small portion of CPU resources. Finally, the differences in memory consumption are negligible and can be considered measurement noise.

\section{Conclusion and future work} 

Aiming to foster greater awareness of energy consumption in AI research, we developed a software toolkit \textsf{AIMeter}, which enables measuring, analyzing, and visualizing energy and carbon footprint of AI workloads. \textsf{AIMeter} also forms a solid foundation for AI energy predicting models. Future work will focus on achieving task-level granularity (e.g., attributing energy consumption to individual processes running concurrently on shared hardware) and extending compatibility beyond NVIDIA GPUs.
\nop{
About the future direction, we plan to extend \textsf{AIMeter}'s capabilities in following aspects:
\begin{itemize}
    \item \textbf{Enhancing Granularity:} Future work will focus on achieving task-level granularity, attributing energy consumption to individual processes running concurrently on shared hardware.
    \item \textbf{Broadening Hardware Support:} Another key direction is extending compatibility beyond NVIDIA GPUs.
\end{itemize}
}

We hope this work contributes meaningfully to the advancement of Green AI principles, inspires further research into sustainable AI practices, and supports efforts to reduce the carbon footprint of modern AI systems.

\bibliographystyle{unsrt}
\bibliography{main} 

\appendix
\section{Metric Explanation}
\label{sec:metrics_appendix}

Table~\ref{tab:gpu_metrics} details the collected GPU performance metrics, categorized by their primary function.

\begin{table*}[ht]
\centering
\small 
\renewcommand{\arraystretch}{1} 
\setlength{\baselineskip}{1em}  
\setlength{\itemsep}{0.3em}  
\caption{GPU Performance Metrics}
\label{tab:gpu_metrics}
\begin{tabular}{|p{3.5cm}|p{13cm}|}
\hline
\rowcolor{gray!30} \textbf{Section} & \textbf{Metric and Description} \\
\hline
\textbf{Energy Section} & 
\begin{itemize}
    \item \textbf{power.draw [W]}: Current real-time power consumption of the GPU, in Watts (W).
    \item \textbf{temperature.gpu}: Current temperature of the main GPU core.
    \item \textbf{cpu\_power}: Current real-time power consumption of the CPU, in Watts (W).
    \item \textbf{dram\_power}: Current real-time power consumption of the DRAM, in Watts (W).
\end{itemize} \\
\hline
\textbf{Compute Section} & 
\begin{itemize}
    \item \textbf{utilization.gpu [\%]}: Percentage of time over the past sample period during which one or more kernels were executing on the GPU's Streaming Multiprocessors (SMs).
    \item \textbf{sm\_active}: Percentage of time and quantity the Streaming Multiprocessors (SMs) were active (executing instructions).
    \item \textbf{sm\_occupancy}: Ratio of active warps on a Streaming Multiprocessor (SM) to the maximum number of warps supported by the SM.
    \item \textbf{tensor\_active}: Percentage of time and quantity the Tensor Cores (used for accelerating AI computations) were active.
    \item \textbf{fp64\_active}: Percentage of time the GPU's double-precision (FP64) units were active.
    \item \textbf{fp32\_active}: Percentage of time the GPU's single-precision (FP32) units were active.
    \item \textbf{fp16\_active}: Percentage of time the GPU's half-precision (FP16) units were active.
    \item \textbf{clocks.current.graphics [MHz]}: Current clock frequency of the GPU's graphics/shader cores, in Megahertz (MHz).
    \item \textbf{clocks.current.sm [MHz]}: Current clock frequency of the GPU's Streaming Multiprocessors (SMs), in Megahertz (MHz).
\end{itemize} \\
\hline
\textbf{Memory Section} & 
\begin{itemize}
    \item \textbf{utilization.memory [\%]}: Percentage of time over the past sample period during which the GPU's memory interface was busy.
    \item \textbf{dram\_active}: The proportion of cycles the interface was actively transferring data, reflecting bandwidth usage efficiency (Analogy: Consider utilization.memory as the time a kitchen pantry door is open, and dram\_active as the time a chef's hands are actually carrying ingredients from it. The door might be open longer than ingredients are being moved.).
    \item \textbf{usage.memory [\%]}: Percentage of total available GPU memory that is currently allocated or used.
    \item \textbf{temperature.memory}: Current temperature of the GPU memory modules.
    \item \textbf{clocks.current.memory [MHz]}: Current clock frequency of the GPU memory, in Megahertz (MHz).
\end{itemize} \\
\hline
\textbf{Communication Section} & 
\begin{itemize}
    \item \textbf{pcie.link.gen.current}: Current generation of the PCIe link, determining the maximum theoretical transfer rate.
    \item \textbf{pcie.link.width.current}: Current number of active PCIe lanes used by the link.
    \item \textbf{pcie\_tx\_bytes}: Total number of bytes transmitted from the GPU to the host via the PCIe bus.
    \item \textbf{pcie\_rx\_bytes}: Total number of bytes received by the GPU from the host via the PCIe bus.
    \item \textbf{nvlink\_tx\_bytes}: Total number of bytes transmitted via the NVLink high-speed interconnect.
    \item \textbf{nvlink\_rx\_bytes}: Total number of bytes received via the NVLink high-speed interconnect.
\end{itemize} \\
\hline
\textbf{System Section} & 
\begin{itemize}
    \item \textbf{cpu\_usage}: Percentage of the host system's CPU utilization.
    \item \textbf{dram\_usage}: Percentage of the host system's main memory (RAM/DRAM) currently in use.
\end{itemize} \\
\hline
\end{tabular}
\renewcommand{\arraystretch}{1} 
\end{table*}

\section{Demonstration Examples}
\label{demonstration}
\begin{figure*}[htbp]
    \centering
    \includegraphics[width=1\linewidth]{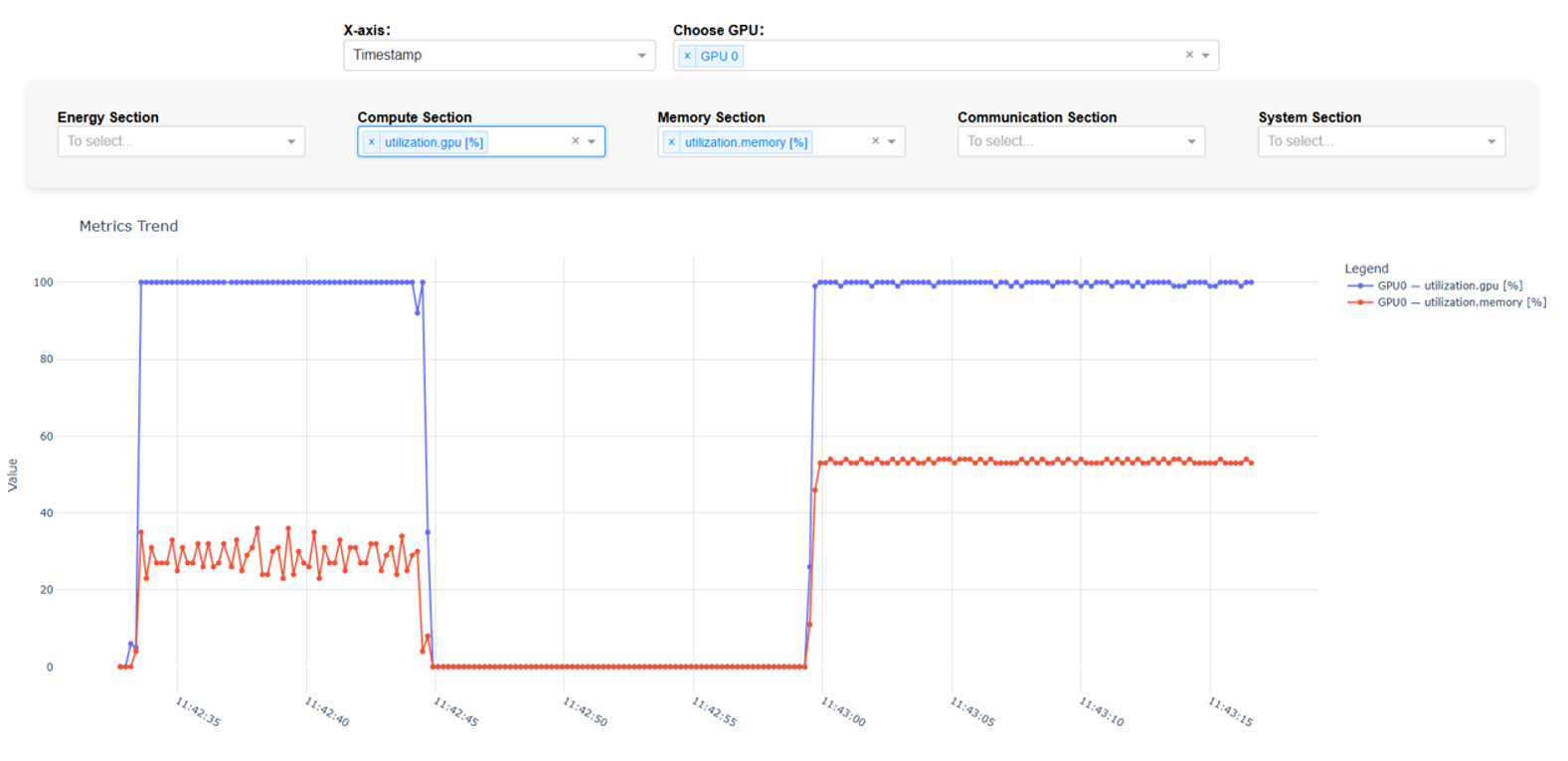}
    \caption{Plotting of GPU activities.}
    \label{fig:plotting}
    \vspace{2cm}
\end{figure*}

\begin{figure*}[htbp]
    \centering
    \includegraphics[width=0.9\linewidth]{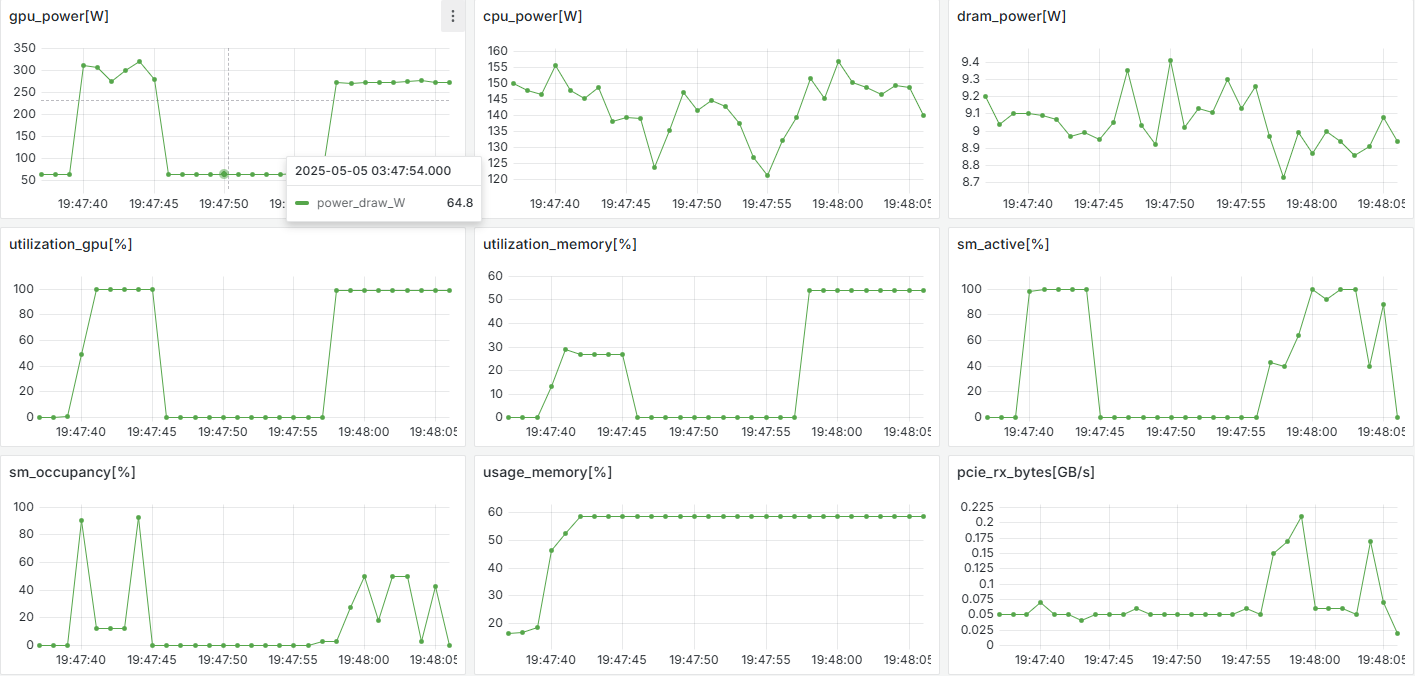}
    \caption{Real-time monitoring dashboard.}
    \label{fig:dashboard}
\end{figure*}

Tables~\ref{tab:combined_perf_metrics},~\ref{tab:tabular_data}, Figs.~\ref{fig:plotting}, and~\ref{fig:dashboard} present examples of the four data demonstration approaches.


\begin{table*}[htbp]
\small
\centering
\caption{Execution Report}
\label{tab:combined_perf_metrics}
\renewcommand{\arraystretch}{1} 
\begin{tabular}{>{\centering\arraybackslash}m{2.6cm} >{\centering\arraybackslash}m{4.6cm}
                >{\centering\arraybackslash}m{2.2cm}
                >{\centering\arraybackslash}m{2.2cm} 
                >{\centering\arraybackslash}m{2.2cm} 
                >{\centering\arraybackslash}m{2.2cm}} 
\toprule
\multicolumn{2}{c}{\textbf{Overall Performance Metric}} & \textbf{Value} & \textbf{} & \textbf{} & \textbf{} \\
\midrule
\multicolumn{2}{c}{Total Time [s]}                     & 43.00 & & & \\
\multicolumn{2}{c}{CPU Energy [J]}                     & 6348.88 & & & \\
\multicolumn{2}{c}{DRAM Energy [J]}                    & 388.74 & & & \\
\multicolumn{2}{c}{GPU 0 Energy [J]}                   & 8918.24 & & & \\
\multicolumn{2}{c}{Total Energy [J]}                   & 15655.86 & & & \\
\multicolumn{2}{c}{Carbon Emissions [kg CO$_2$eq]}     & 0.0020 & & & \\
\midrule

\textbf{Component} & \textbf{Metric} & \textbf{Avg} & \textbf{Max} & \textbf{Min} & \textbf{Mode} \\
\midrule
\multirow{2}{*}{CPU} & Usage [\%] & 3.67 & 6.20 & 2.20 & 4.00 \\
                     & Power [W]  & 147.91 & 160.26 & 123.73 & 123.73 \\
\addlinespace
\multirow{2}{*}{DRAM} & Usage [\%] & 4.10 & 4.10 & 3.90 & 4.10 \\
                      & Power [W]  & 9.05 & 9.43 & 8.77 & 9.06 \\
\midrule

\multicolumn{6}{c}{\textbf{GPU 0 Detailed Statistics (NVIDIA A800 80GB PCIe)}} \\
\midrule
\textbf{Category} & \textbf{Metric} & \textbf{Avg} & \textbf{Max} & \textbf{Min} & \textbf{Mode} \\
\midrule
\multirow{2}{*}{Energy} & Power Draw [W] & 204.12 & 315.58 & 62.89 & 65.86 \\
                        & GPU Temp. [°C] & 43.09 & 51.00 & 32.00 & 38.00 \\
\addlinespace
\multirow{9}{*}{Compute} & GPU Utilization [\%] & 63.55 & 100.00 & 0.00 & 100.00 \\
                         & Graphics Clock [MHz] & 1354.43 & 1410.00 & 1125.00 & 1410.00 \\
                         & SM Clock [MHz] & 1354.43 & 1410.00 & 1125.00 & 1410.00 \\
                         & SM Active [\%] & 59.91 & 100.00 & 0.00 & 100.00 \\
                         & SM Occupancy [\%] & 20.16 & 92.30 & 0.00 & 0.00 \\
                         & Tensor Active [\%] & 19.91 & 93.70 & 0.00 & 0.00 \\
                         & FP64 Active [\%] & 0.00 & 0.00 & 0.00 & 0.00 \\
                         & FP32 Active [\%] & 1.45 & 31.90 & 0.00 & 0.00 \\
                         & FP16 Active [\%] & 0.13 & 5.10 & 0.00 & 0.00 \\
\addlinespace
\multirow{5}{*}{Memory} & Mem. Utilization [\%] & 27.41 & 54.00 & 0.00 & 0.00 \\
                        & Mem. Temp. [°C] & 45.91 & 55.00 & 35.00 & 41.00 \\
                        & Mem. Clock [MHz] & 1512.00 & 1512.00 & 1512.00 & 1512.00 \\
                        & Mem. Usage Total [\%] & 88.94 & 94.56 & 17.23 & 90.02 \\
                        & DRAM Active (Cycles) [\%] & 18.40 & 94.60 & 0.00 & 0.00 \\
\addlinespace
\multirow{6}{*}{Communication} & PCIe Link Gen & 4.00 & 4.00 & 4.00 & 4.00 \\
                               & PCIe Width & 16.00 & 16.00 & 16.00 & 16.00 \\
                               & PCIe TX [GB/s] & 0.11 & 0.14 & 0.07 & 0.10 \\
                               & PCIe RX [GB/s] & 0.05 & 0.13 & 0.04 & 0.05 \\
                               & NVLink TX [GB/s] & 0.00 & 0.00 & 0.00 & 0.00 \\
                               & NVLink RX [GB/s] & 0.00 & 0.00 & 0.00 & 0.00 \\
\midrule[\heavyrulewidth] 

\multicolumn{6}{c}{\textbf{Top Positively and Negatively Correlated Metric Pairs}} \\
\midrule
\multicolumn{6}{c}{%
  \centering 
  \renewcommand{\arraystretch}{1.1} 
  \begin{tabular}{
    >{\centering\arraybackslash}m{3.0cm} 
    >{\centering\arraybackslash}m{3.5cm} 
    >{\centering\arraybackslash}m{1.5cm} 
    >{\centering\arraybackslash}m{3.0cm} 
    >{\centering\arraybackslash}m{3.5cm} 
    >{\centering\arraybackslash}m{1.5cm} 
  }
  \textbf{Metric A} & \textbf{Metric B} & \textbf{Coeff.} & \textbf{Metric A} & \textbf{Metric B} & \textbf{Coeff.} \\
  \midrule
  SM Clock [MHz] & Graphics Clock [MHz] & 1.000 & SM Clock [MHz] & Tensor Active [\%] & -0.844 \\
  GPU Utilization [\%] & Power Draw [W] & 0.966 & Tensor Active [\%] & Graphics Clock [MHz] & -0.844 \\
  SM Active [\%] & Power Draw [W] & 0.948 & SM Active [\%] & PCIe TX [GB/s] & -0.807 \\
  GPU Utilization [\%] & SM Active [\%] & 0.938 & PCIe TX [GB/s] & Power Draw [W] & -0.768 \\
  Mem. Temp. [°C] & GPU Temp. [°C] & 0.924 & GPU Utilization [\%] & PCIe TX [GB/s] & -0.650 \\
  \end{tabular}%
} \\
\bottomrule
\end{tabular}
\end{table*}


\begin{table*}[htbp]
    \centering
    \caption{Tabular Data}
    \includegraphics[width=1\linewidth]{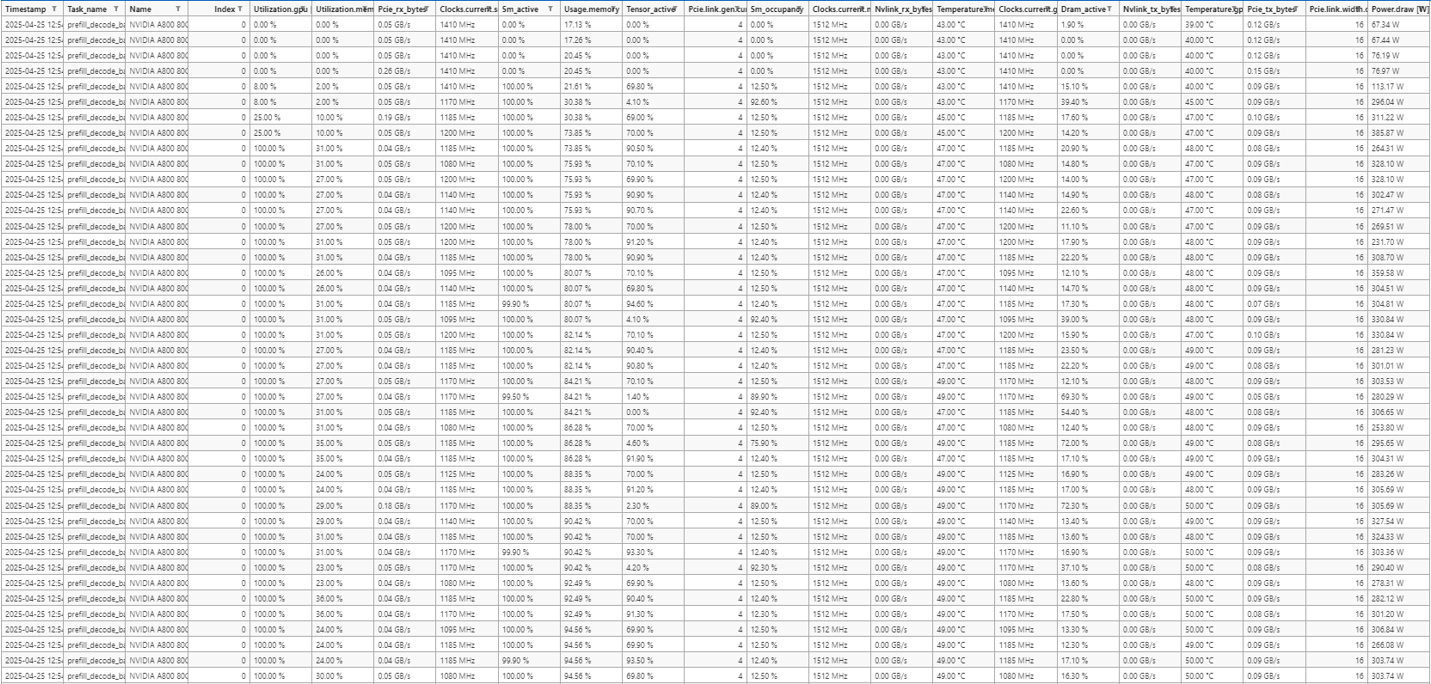}
    \label{tab:tabular_data}
\end{table*}

\end{document}